\documentclass[12pt]{iopart}
\usepackage{graphicx}
\usepackage{iopams}
\usepackage{setstack}
\usepackage{bm}
\usepackage{amsfonts}
\usepackage{amssymb}
\eqnobysec

\begin{document}

\title[Survival and percolation probabilities]{Survival and percolation probabilities\\ in the field theory of growth models}

\author{Hans-Karl Janssen}

\address{Institut f\"{u}r Theoretische Physik III, Heinrich-Heine-Universit\"{a}t,\\
40225 D\"{u}sseldorf, Germany }

\ead{janssen@thphy.uni-duesseldorf.de}

\begin{abstract}
Survival and percolation probabilities are most important quantities in the
theory and in the application of growth models with spreading. We construct
field theoretical expressions for these probabilities which are feasible for
perturbation expansions. The outstanding role of the absorbing noise is
stressed to obtain survival probabilities monotonic decreasing with time. We
briefly consider some fundamental growth models equipped with absorbing noise
which are representations of known universality classes of spreading
phenomena. The critical scaling properties of their survival and percolation
probabilities are stated. In an appendix we consider shortly the renormalized
field theory of compact directed percolation.
\end{abstract}





\section{Introduction}

\label{intro} The investigation of the formation and properties of random
structures has been an exciting topic in statistical physics for many years
(see e.g.\ \cite{StAh94,BuHa95/96}). Of particular interest are random
structures which are formed by local rules. These processes can often be
expressed in the language of population growth. Stochastic processes of this
type describe the essential features of a vast number of growth phenomena of
populations near their extinction threshold, and are relevant to a wide range
of models in physics, chemistry, biology, and sociology
\cite{Hi01,Mo77,Ba85,Li85,Mur89}. The transition between survival and
extinction of such a growing population constitutes a nonequilibrium
continuous phase transition phenomenon and is characterized by universal
scaling laws. The extinct states without active individuals (interacting
particles) are stochastically absorbing, i.e. in a finite system with a finite
number of degrees of freedom an extinct inactive state is reached with
probability one at a finite time. Beyond the critical transition point, where
the replication rate is greater than a critical value, the systems are in the
active state. The corresponding mean extinction time is practically infinite
in systems with a great number of degrees of freedom. To find active steady
states it is therefore allowed to take the infinite volume limit, the
so-called thermodynamic limit, before the infinite time limit is done. Then,
the probability $P_{\infty}$ that an active state, created by a localized seed
is still alive after an infinite time is greater than zero beyond the critical
point. Hence, this percolation probability serves besides the density of the
active particles in the steady state, $\rho_{st}$, as an order parameter for
growth processes. On the other hand, in the absorbing phase, the phase below
the critical replication rate, each active configuration dies after a finite
time. Thus $P_{\infty}$ is zero here and the steady states are inactive.
However, an interesting observable in both phases is the survival probability
$P(t)$. It is the probability that a process started from a localized seed,
e.g. an active particle at the origin of space and time, is still active at
time $t$. Of course, we have generally $P(t)\rightarrow P_{\infty}$ for
$t\rightarrow\infty$.

Near a critical point one is typically interested in the asymptotic universal
long-wavelength, long-time behavior of the observables. Besides the existing
vast number of simulation methods and results (see e.g.\ \cite{Hi01}), a
variety of analytical tools were developed for growth processes in the past
decades. The method which introduces the continuum limit from the beginning
and exploits the renormalization group to obtain universal properties is the
renormalized stochastic field theory \cite{Ja76,DeDo76,Ja92,JaTa04}. At the
heart of this method is the calculation of time dependent mean values,
correlation and response functions of the observables by using path integrals.
Basic ingredient is a generalized probability measure $\exp(-\mathcal{J})$,
where the dynamic functional $\mathcal{J}$ (nowadays also called a response
functional) is a functional of time and space dependent fields. The
restriction of $\mathcal{J}$ to the only relevant terms in an expansion of the
slowly varying fields defines a renormalizable field theory. This theory
represents then a full universality class of systems with common asymptotic properties.

Although the field theory of growth processes are based on a probability
measure, the direct calculation of the survival probability $P(t)$ and the
percolation probability $P_{\infty}$ for growth processes by field theoretic
means was not well developed in the past. There exists an attempt of
Mu\~{n}oz, Grinstein, and Tu \cite{MuGrTu97} (henceforth cited as MGT) to devise a practical method to determine correctly the scaling properties of the survival probability.
However, these authors erroneously assert: \textquotedblleft In field theories
with continuous variables \ldots\ the absorbing state is a set of measure zero
in phase space and so can never actually be reached in finite time. Thus
$P(t)$ is strictly equal to unity for all $t$, so the concept of the survival
probability has no utility.\textquotedblright, an argument that resembles
Zenon's paradox on Achilles and the tortoise \cite{Ho79}. They introduce
therefore another quantity as a substitute for $P(t)$ with the correct scaling
properties. But their argumentation seems to rest solely on a naive
interpretation of the Langevin equation as an ordinary differential equation
for a continuously differentiable function, and of a stationary solution of
the corresponding Fokker-Planck equation for the probability density of the
stochastic process $\phi$ in zero spatial dimension. However, this stationary
solution has a non-integrable singularity at $\phi=0$, and is therefore not
normalizable. Hence, it cannot represent any features of the correct
probability density near the absorbing state $\phi=0$. MGT, without 
considering any boundary condition of the time dependent solution of 
the Fokker-Planck equation, conclude: \textquotedblleft An arbitrary 
initial probability distribution therefore evolves in time toward
a distribution weighted at values $\phi$ lying progressively closer to zero.
Note, however, that $\phi$ cannot actually achieve the value $0$ in finite
time.\textquotedblright\ However, to reveal such properties of the time
dependent probability distribution one has to ascertain analytically the
behavior of the solutions of the Fokker-Planck equation near and at the
absorptive state. But this was not done by MGT.
Hence, their arguments are far of being convincing. We will show that the
validity of their conclusion depends heavily on the form of the underlying
noise. For a noise called absorptive in the following, the absorbing state has
a $\delta$-probability measure in the continuous phase space. This measure
accumulates more and more mass during the course of time. Consequently, the
survival probability $P(t)$ is a monotonically decreasing function also in a
continuous phase space. In contrast, for the so-called multiplicative noise
\cite{Mu98} the assertions MGT are correct, and $P(t)=1$ for all finite times.

In this paper we develop expressions for survival and percolation
probabilities of several reaction-diffusion processes with absorbing states in
non-zero spatial dimensions from which the calculation of these quantities by
field-theoretical methods can be performed. We show the feasibility of our
expressions for the calculation of nontrivial probabilities in a mean field
(saddle point) approximation that becomes correct for spatial extended systems
above their higher critical dimension.

The outline of our paper is as follows. In section~II we present a simple toy
model that shows the fundamental difference between absorptive and
multiplicative noise. It provides a nontrivial survival probability in the
absorbing case. Section~III reviews well known field theoretic results for
directed percolation (DP) and its properties under the duality transformation.
In section~IV we derive the field theoretic expressions of the survival and
the percolation probability for DP. Asymptotic scaling forms are found that
are related to the mean particle density of the dual process. In section~V we
calculate the probabilities in a saddle point approximation. In section~VI we
study a variant of DP equipped with multiplicative noise. We show that in this
case the survival probability is equal to one for all finite times. We compare
in this section our result for the survival probability with the expression
introduced by MGT. In Sec.~VII we apply our results to the field
theories of various fundamental growth processes. A short epilogue is given in
section~VIII. An appendix contains a brief presentation of the renormalized
field theory of compact directed percolation.

\section{A toy model}

For the ease of the reader we re-consider in this section a well known example
of a continuous growth process for which survival probabilities can be easily
discussed in a concise way. Most of this material can be found in textbooks
\cite{Fe52,BR60,Gar85,vKa92}, but it does not seem much appreciated by those
working in this research field. The stochastic version of the logistic or
Malthus-Pearl-Verhulst equation, $\dot{x}=-r(x)x+\zeta$ with $r(x)=(\tau+gx)$,
describes the growth of a population size $x(t)\geq0$. Equipped with
absorptive noise $\zeta$, it provides a zero-dimensional toy-model of directed
percolation. The model is also useful for the mean field description of the
evolution of a homogeneously distributed population in a spatial extended
system with volume $V$. In this case one defines the particle density by
$n(t)=x(t)/V$. The coupling constant scales with the volume as $g\sim1/V$. The
Gaussian random noise $\zeta(t)$ is characterized by correlations
$\overline{\zeta(t)\zeta(t^{\prime})}=2x(t)^{\alpha}\delta(t-t^{\prime})$. For
physical growth processes these correlations as functions of $x$, $t$, and
$t^{\prime}$ result from the elimination of many fast microscopic degrees of
freedom in local equilibrium with the population size $x$. Therefore, they
should be analytic functions in $x$ and zero if $x=0$. Hence, in the limit
where microscopic correlation times go to zero on the slow macroscopic
time-scale, we typically find for population processes $\alpha=1$ (absorptive
noise) or $\alpha=2$ (multiplicative noise) in a small $x$ expansion
neglecting higher orders. The nowadays popular terminology \textquotedblleft
multiplicative noise\textquotedblleft\ stems from the fact that in the
before-mentioned time limit, the noise may be formally written as a product of
an analytic function of $x$ and of white noise. In any event, it is essential
that the noise is interpreted in the It\={o}-sense \cite{Gar85,vKa92}. A very
qualitative glance on the logistic Langevin-equation shows that the random
impulses resulting from the noise scale like $x^{\alpha/2}$. Thus, for small
$x$, these impulses can easily surmount the distance $x$ to the absorbing
state $x=0$ if $\alpha<2$. We expect that this is the condition for which a
non-zero extinction probability $P_{0}(t)=1-P(t)$ for finite times can be defined.

Nevertheless, a more satisfactory tool for the analytical description of this
stochastic process is given by the corresponding Fokker-Planck equation for
the probability density $p(x,t)$ to find a population with $x\geq0$ at time
$t$:
\begin{equation}
\frac{\partial p(x,t)}{\partial t}=\frac{\partial\lbrack r(x)xp(x,t)]}
{\partial x}+\frac{\partial^{2}[x^{\alpha}p(x,t)]}{\partial x^{2}}.
\label{FP-Eq} 
\end{equation}
The Fokker-Planck equation is also known as the forward Kolmogorov equation
for the fundamental probability density $p(x,t)=w(x,t|x^{\prime},t^{\prime})$
with $t\geq t^{\prime}$ and the initial condition $w(x,t^{\prime}|x^{\prime
},t^{\prime})=\delta(x-x^{\prime})$ with $0<x^{\prime}<\infty$. As a second
order partial differential equation it needs in general two boundary
conditions at $x=0$ and $x=\infty$. Starting in the open interval $(0,\infty
)$, it may be that the stochastic process $x(t)$ does never takes on the value
$x=0$ or $x=\infty$ in finite times. In this case the corresponding boundary
is called natural, and no boundary condition at this value has to be imposed.
In contrast, if the process $x(t)$ has positive probability of taking the
value $x=0$ or $x=\infty$, the corresponding boundary is called accessible.
For accessible boundaries there are two possible cases:

\begin{enumerate}
\item Exit: the drift towards the boundary is such that the boundary
automatically acts as an absorbing barrier, and no boundary condition can be imposed.

\item Regular: the process behaves like a classical diffusion process near the
boundary, and various boundary conditions can be imposed.
\end{enumerate}

\noindent Hence, for the existence of an unbiased nontrivial extinction
probability, the absorbing state at $x=0$ should be an exit boundary.

According to the scheme of Feller \cite{Fe52}, the classification of
boundaries depends on the Lebesgue integrability of the function
\begin{eqnarray}
F(x)  &  =\exp\Big\{-\int_{x_{0}}^{x}\rmd y\,r(y)y^{1-\alpha}\Big\}\nonumber\\
&  =\exp\Big\{\frac{\tau\bigl(x_{0}^{2-\alpha}-x^{2-\alpha}\bigr)}{2-\alpha
}+\frac{g\bigl(x_{0}^{3-\alpha}-x^{3-\alpha}\bigr)}{3-\alpha}\Big\}\,,
\label{Phi} 
\end{eqnarray}
where $x_{0}\in(0,\infty)$. The following functions are defined besides
$F(x)$:
\numparts
\begin{eqnarray}
\label{HilfsFu}
G(x)  &  =\frac{F(x)}{x^{\alpha}}\,,\label{G}\\
H(x)  &  =F(x)^{-1}\int_{x_{0}}^{x}\rmd y\,G(y)\,. \label{H} 
\end{eqnarray}
\endnumparts
Note that the stationary solution of the Fokker-Planck equation (\ref{FP-Eq}),
$G(x)$, is a stationary probability distribution only if it is normalizable.
We denote with $f(x)\in\mathcal{L}(0)$ or $f(x)\in\mathcal{L}(\infty)$ a
function that is Lebesgue integrable on the interval $(0,x_{0})$ or
$(x_{0},\infty)$, respectively. Then Feller's classification criteria are:
\begin{enumerate}
\item The boundary $b$ is regular if $F(x)^{-1}\in\mathcal{L}(b)$,
$G(x)\in\mathcal{L}(b)$, and $H(x)\in\mathcal{L}(b)$;

\item The boundary $b$ is exit if $F(x)^{-1}\in\mathcal{L}(b)$, 
$G(x)\notin\mathcal{L}(b)$ and $H(x)\in\mathcal{L}(b)$;

\item The boundary $b$ is natural in the other cases.
\end{enumerate}

\noindent Applying these criteria one finds easily that for $\alpha<3$ the
boundary at infinity is natural, whereas the boundary at zero is regular for
$\alpha<1$, exit for $1\leq\alpha<2$, and natural for $2\leq\alpha$
(attractive for $\tau>-1$, repulsive for $\tau\leq-1$). It is now easily seen
that for $1\leq\alpha<2$ the Fokker-Planck equation (\ref{FP-Eq}) is
transformed back to the case $\alpha=1$ by the setting $y=x^{2-\alpha}$ and
the replacement $r(x)\rightarrow\tilde{r}(y)=r(x)+(\alpha-1)/(2-\alpha)$.
Hence, in the following we concentrate on the absorptive noise $\alpha=1$ and
to the multiplicative noise $\alpha=2$. In both cases no additional boundary
conditions are needed for the solution of the initial problem.

On the other hand we find from above that there is a fundamental difference
between these two cases: for absorptive noise there is a non-zero probability
for the stochastic process to reach the absorbing state at $x=0$ in finite
times, whereas for multiplicative noise the probability is zero that the
process reaches $x=0$ in finite times. This different behavior is reflected in
the evolution of the probability distribution in time: for absorptive noise
the probability density shows a $\delta$-peak directly at the absorbing
boundary which accumulates more and more mass in the course of time. For
multiplicative noise in the case $\tau>-1$ the probability distribution
$p(x,t)$ represents a peak, which moves to the left, becomes smaller and
smaller in extension, and accumulates all probability near the absorbing state
$x=0$ without reaching this state itself ($p(0,t)=0$ for $t<\infty$). Only in
the infinite time limit we get $\lim_{t\rightarrow\infty}p(x,t)=\delta(x)$ and
the stochastic process is ultimately absorbed. For $\tau<-1$ the probability
distribution with $p(0,t)=0$ for $t<\infty$ converges to the stationary
distribution $\lim_{t\rightarrow\infty}p(x,t)\propto G(x)$, equation (\ref{G}).
However, note that the limits $\lim_{t\rightarrow\infty}$ and $\lim
_{x\rightarrow0}$ cannot be interchanged \cite{GraSche82}.

The fundamentally different behavior of the distributions in time can be made
fully explicit in the linear case $r(x)=\tau$ \cite{BR60,vKa92,GraSche82}. If
the noise is absorptive, $\alpha=1$, the solution of the Fokker-Planck
equation (\ref{FP-Eq}) is found easily by a Laplace transform to
\numparts
\begin{eqnarray} \label{abFPSol}
p(x,t|x_{0})  &  =\delta(x)P_{0}(t,x_{0})+\theta(x)\bar{p}(x,t|x_{0}
)\,,\label{p}\\
\bar{p}(x,t|x_{0})  &  =h(t)\sqrt{\bar{x}(t)/x}\,I_{1}\bigl(2h(t)\sqrt{\bar
{x}(t)x}\bigr)\exp\bigl[-h(t)(\bar{x}(t)+x)\bigr]\,,\\
P_{0}(t,x_{0})  &  =\exp\bigl[-h(t)\bar{x}(t)\bigr]\,. \label{ExtP}
\end{eqnarray}
$I_{1}$ is the modified Bessel function of first order, and we have defined
\begin{eqnarray}
\bar{x}(t)  &  =x_{0}\exp(-\tau t)\,,\\
h(t)  &  =\frac{\tau}{1-\exp(-\tau t)}\,.
\end{eqnarray}
\endnumparts
Note that the extinction probability (\ref{ExtP}) increases in the course of
time monotonically from $0$ to $1$ (or $\exp(-\left\vert \tau\right\vert
x_{0})$ if $\tau<0$, whereby $P_{\infty}(x_{0})=\lim_{t\rightarrow\infty
}(1-P_{0}(t,x_{0}))=1-\exp(-\left\vert \tau\right\vert x_{0})$ is a kind of
percolation probability). Hence, it is explicitly shown that the assertion 
\textquotedblleft the stochastic process cannot actually achieve the value $0$
in finite time\textquotedblright\ of MGT is wrong in the case of absorptive noise.

The behavior of the probability distribution is qualitatively different in the
case of multiplicative noise, $\alpha=2$. Then, the Fokker-Planck equation
(\ref{FP-Eq}) is solved by the substitution $x=\exp y$ that transforms the
stochastic process to simple diffusion with drift. One obtains
\begin{eqnarray}
p(x,t|x_{0})  &  =\bigl(\sqrt{4\pi t}x\bigr)^{-1}\exp\Big(-\bigl[\ln x/x_{0}+(\tau+1)t\bigr]^{2}/4t\Big)\,.
\end{eqnarray}
Indeed, here $p(0,t|x_{0})=0$ and $P_{0}(t,x_{0})=0$ for all finite times, and
the assertion of MGT is correct.

The fundamental probability density $w(x,t|x^{\prime},t^{\prime})$ must also
be a solution of the backward Kolmogorov equation
\begin{equation}
\frac{\partial w(x,t|x^{\prime},t^{\prime})}{\partial t^{\prime}}=r(x^{\prime
})x^{\prime}\frac{\partial w(x,t|x^{\prime},t^{\prime})}{\partial x^{\prime}
}-x^{\prime\alpha}\frac{\partial^{2}w(x,t|x^{\prime},t^{\prime})}{\partial
x^{\prime2}}\,. \label{bwK-Eq} 
\end{equation}
With the Ansatz $w(x,t|x^{\prime},t^{\prime})=\delta(x)P_{0}(t-t^{\prime
},x^{\prime})+\theta(x)\bar{p}(x,t|x^{\prime},t^{\prime})$ one formally
extends the range of the variable $x$ in (\ref{FP-Eq}) to the open interval
$(-\infty,\infty)$. Then $\int_{-\infty}^{\infty}\rmd x\,p(x,t|x^{\prime
},t^{\prime})=P_{0}(t-t^{\prime},x^{\prime})+\int_{0}^{\infty}\rmd x\,\bar
{p}(x,t|x^{\prime},t^{\prime})=1$ where $\bar{p}(x,t|x^{\prime},t^{\prime})$
fulfills the original Kolmogorov equations. For $P_{0}(t,x^{\prime})$ one gets
the ordinary differential equation
\begin{eqnarray}
\fl \qquad\qquad \dot{P}_{0}(t-t^{\prime},x^{\prime}) =-\frac{d}{dt}\int_{0}^{\infty
}\rmd x\,\bar{p}(x,t|x^{\prime},t^{\prime})=\alpha\lim_{x\rightarrow0}
\bigl(x^{\alpha-1}\bar{p}(x,t|x^{\prime},t^{\prime})\bigr)\,.
\end{eqnarray}
$\dot{P}_{0}$ is zero for multiplicative noise but equal to $\bar
{p}(0,t|x^{\prime},t^{\prime})\,$for absorptive noise. From the backward
equation (\ref{bwK-Eq}) on derives the differential equation of the survival
probability $P(t,x^{\prime})=1-P_{0}(t|x^{\prime},t^{\prime}=0)$ starting from
a state with population size $x^{\prime}>0$ at $t^{\prime}=0$. Using the time
translation invariance of the process, the time evolution follows from the
backward Kolmogorov equation (\ref{bwK-Eq}) as
\begin{equation}
\frac{\partial P(t,x)}{\partial t}=-r(x)x\frac{\partial P(t,x)}{\partial
x}+x^{\alpha}\frac{\partial^{2}P(t,x)}{\partial x^{2}} \label{bwFP-Eq} 
\end{equation}
with $P(0,x)=1$ for $x>0$. Using this equation the mean time $T(x)=\int
_{0}^{\infty}\rmd t\,P(x,t)$ to reach the absorbing boundary at $x=0$ from a state
$x>0$ is easily calculated
\begin{equation}
T(x)=\int_{0}^{x}\rmd y\int_{0}^{\infty}\frac{\rmd s}{(s+y)^{\alpha}}\exp
\Bigl(-\int_{y}^{s+y}\rmd z\,z^{1-\alpha}r(z)\Bigr)\,.
\end{equation}
With $r(x)=\tau+gx$ one obtains in the case of multiplicative noise
\begin{equation}
T(x)=\int_{0}^{x}\rmd y\,y^{\tau}\int_{0}^{\infty}\frac{\rmd s}{(s+y)^{2+\tau}}
\exp\bigl(-gs\bigr)\,.
\end{equation}
This expression diverges at the lower integration boundary, $y=0$,
irrespectively of the sign of $\tau$ for all $x>0$. Hence, as we already know,
the stochastic process cannot reach the absorbing state in finite time.

In the case of absorptive noise we obtain
\begin{equation}
T(x)=\int_{0}^{x}\rmd y\int_{0}^{\infty}\frac{\rmd s}{s+y}\exp\bigl[-gs^{2}
/2-(\tau+gy)s\bigr]. \label{ex_time} 
\end{equation}
This expression is finite for all $x>0$ and shows again that the extinction
probability to reach the absorbing state is nonzero even for finite times.
From the mean field point of view, starting with a homogeneous state in the
volume $V$, we have $x\sim V$ and $g\sim1/V$. Consequently, for $\tau>0$ the
mean extinction time scales with the volume as $T(x)\sim\tau^{-1}\left\vert
\ln x\right\vert $, whereas for $\tau<0$ we have an exponential increase $\ln
T(x)\sim\tau^{2}/g$. This profound difference between the absorbing and the
active phase of population growth is also expected in extended but finite systems.

Our discussion of known results from the analysis of continuous stochastic
processes has shown that the behavior of the survival probability in time
depends crucially on the type of the noise. It is shown that there exists a
non-trivial survival probability in a theory with absorptive noise also in a
phase space of a continuous variable. Thus, the toy model does not provide any
argument against the existence of a non-trivial survival probability in the
case of field theories of growth processes in spatial dimensions greater than
zero. Indeed, in the next chapters we directly construct calculable
expressions of survival and percolation probabilities for such field theories
in terms of path integrals.

\section{Directed percolation}

In order to be consistent with later parts of the article, we will concisely
present well known material on critical stochastic processes with absorbing
states in this section. We will concentrate on directed percolation (DP) as
the typical growth model \cite{BroHa57,GraSu78,CaSu80,Ob80,Ja81,Ja01,Ki83}.
For an comprehensive overview see \cite{Hi01,JaTa04}.

From the onset we use a mesoscopic picture in which all microscopic length-
and time-scales are considered to be very short. We are interested in the
asymptotic properties of the growth process near the critical point between
the absorbing phase and the steady active phase. Thus, we take a continuum
approach with the density of the active particles, $n\left(\mathbf{x} ,t\right)$, 
as the stochastic variables. We remove all irrelevant
contributions to the equations of motion. The growth process might be
represented by a Langevin equation (in the It\={o} sense), which is
constructed in accordance with the principle of the existence of an absorbing
state as \cite{Ja81,Ja85,Ja01}
\begin{equation}
\partial_{t}n=\lambda\nabla^{2}n-R\left[  n\right]  n+Q+\zeta, \label{LangGl} 
\end{equation}
where $\zeta\left(  \mathbf{x},t\right)  $ denotes the noise which has to
vanish if $n\left(  \mathbf{x},t\right)  =0$. The reaction rate $R\left[
n\right]  $ models birth and death of particles, as well as the saturation of
the population. In addition we introduce an external particle source
$Q(\mathbf{x},t)$. The Langevin source $\zeta(\mathbf{x},t)$ can be assumed to
be Gaussian with absorbing correlation
\begin{equation}
\overline{\zeta(\mathbf{x},t)\zeta(\mathbf{x}^{\prime},t^{\prime})}=\lambda
g^{\prime\prime}n(\mathbf{x},t)\delta(\mathbf{x}-\mathbf{x}^{\prime})\delta(t-t^{\prime}). \label{LangQu} 
\end{equation}
Diffusive noise is irrelevant in comparison to the reaction noise,
equation (\ref{LangQu}), in the case of DP, but in general indispensable for
processes with multiplicative noise \cite{Ja81,HoTa97}. The extinction rate is
given by
\begin{equation}
R\left[  n\right]  =\lambda\Big(\tau+\frac{g^{\prime}}{2}n\Big)\,.
\label{R_DP} 
\end{equation}
Under the influence of fluctuations, the critical point is found at a value
$\tau=\tau_{c}<0$. We implicitly renormalize this parameter by $\tau
\rightarrow\tau-\tau_{c}$ so that the critical point is always found at
$\tau=0$.

The field $Q(\mathbf{x},t)\geq0$ of the Langevin equation (\ref{LangGl})
describes the external source distribution of particles. As a special case the
creation of one particle as a seed at the origin $\mathbf{x}=0$ and $t=0$ is
represented by a source term
\begin{equation}
Q(\mathbf{x},t)=\delta(\mathbf{x})\delta(t)\,. \label{AnfQu} 
\end{equation}
On the other hand, the choice $Q=\lambda h=\mathrm{const}$ describes the
creation of particles at a constant rate $Q$ uniform in space and time. In the
following we are interested in the particular source model
\begin{equation}
Q(\mathbf{x},t)=\lambda h+\rho_{0}\delta(t)=\lambda\bigl(h+\rho_{0}
\delta(\lambda t)\bigr)\,, \label{Quelle} 
\end{equation}
where $\rho_{0}$ constitutes a homogeneous particle density as an initial condition.

Dynamic response functionals $\mathcal{J}$ \cite{Ja76,DeDo76,Ja92} based on
Langevin equations of the form (\ref{LangGl}) with noise correlation
(\ref{LangQu}) have the general form \cite{Ja85,Ja81,Ja01}
\begin{equation}
\mathcal{J}[\tilde{n},n]=\int \rmd t\,\rmd ^{d}x\,\biggl\{\lambda\tilde{n} 
\Big(\lambda^{-1}\frac{\partial}{\partial t}+(\tau-\nabla^{2})+\mathcal{V} 
[\tilde{n},n]\Big)n\biggr\}\,. \label{DynFu_n} 
\end{equation}
Here, $\tilde{n}(\mathbf{x},t)$ denotes the response field conjugated to the
particle density $n(\mathbf{x},t)$. For DP we have
\begin{equation}
\mathcal{V}[\tilde{n},n]=\frac{1}{2}\bigl(g^{\prime}n-g^{\prime\prime} 
\tilde{n}\bigr)\,. \label{VDP} 
\end{equation}
The introduction of the source $Q$ leads to the shift
\begin{equation}
\mathcal{J}[\tilde{n},n]\rightarrow\mathcal{J}_{Q}[\tilde{n},n]=\mathcal{J} 
[\tilde{n},n]-(Q,\tilde{n})
\end{equation}
where $(Q,\tilde{n})$ denotes the integral $\int \rmd t\,\rmd^{d}x\,Q(\mathbf{x} 
,t)\tilde{n}(\mathbf{x},t)$. The responses are defined with respect to the
particle source $Q(\mathbf{x},t)\geq0$. They are represented by correlation
functions with the response field $\tilde{n}(\mathbf{x},t)$. Such response and
correlation functions (called Green's functions in the following) can now be
calculated by path integrals with the probability weight $\exp(-\mathcal{J})$.
We use the notation
\begin{equation}
\langle\mathcal{A}[n,\tilde{n}]\rangle=\int\mathcal{D}(\tilde{n} 
,n)\mathcal{A}[n,\tilde{n}]\exp(-\mathcal{J}[\tilde{n},n]),
\end{equation}
where $\mathcal{A}[n,\tilde{n}]$ is any functional of the fields $n,\tilde{n} 
$, and $\mathcal{D}(\tilde{n},n)$ denotes the integral measure which is
properly normalized $\langle1\rangle=1$. In a suitable discretization it reads
$\mathcal{D}(\tilde{n},n)=\prod_{\mathbf{x},t}(d\tilde{n}(\mathbf{x} 
,t)dn(\mathbf{x},t)/2\pi i)$. For practical calculations, the observables
$\mathcal{A}[n,\tilde{n}]$ should be either polynomial or exponential. Then a
diagrammatic perturbation theory can be developed. The integration of the
response fields is done along the imaginary axis, but it can be deformed to
general complex values in finite regions. The integration of the density field
can be formally extended over the full real axis. Initial and final conditions
are supplied by
\begin{equation}
n(\mathbf{x},-\infty)=\tilde{n}(\mathbf{x},\infty)=0 \label{RandB} 
\end{equation}
for all $\mathbf{x}$. For $Q\equiv0$ we have therefore
\begin{equation}
\langle n(\mathbf{x},t)\rangle=\langle\tilde{n}(\mathbf{x},t)\rangle=0
\label{Absorb} 
\end{equation}
for all $\mathbf{x}$ and $t$.

Now we define the duality transformation (known as rapidity reversal in
Reggeon field theory and corresponding to duality in the mathematical theory
of interacting particle systems \cite{Li85}) as
\begin{equation}
\alpha^{-1}\tilde{n}(\mathbf{x},t)=:\tilde{s}(\mathbf{x},t)\longleftrightarrow
-s(\mathbf{x},-t):=-\alpha n(\mathbf{x},-t)\,. \label{DualTr} 
\end{equation}
The free, bilinear, part of the dynamic functional (\ref{DynFu_n}) is
invariant under this transformation for each finite $\alpha$. Note that
$\alpha^{2}$ has the dimension of a spatial volume. Choosing $\alpha
=\sqrt{g^{\prime}/g^{\prime\prime}}$ the interaction part of DP,
$\mathcal{V}[\tilde{n},n]$ (\ref{VDP}), is (up to the time inversion) also
invariant under duality. Of course, $\alpha$ is a redundant variable from the
renormalization group point of view. Such variables should be eliminated
before any renormalization group consideration is applied. Choosing therefore
$s$ and $\tilde{s}$ as the fundamental fields, the invariant coupling is
$g=\sqrt{g^{\prime}g^{\prime\prime}}$, and the fields $s$ and $\tilde{s}$ have
the same scaling dimensions in DP. For general growth models, the duality
transformation defines a dual stochastic process which can be different from
the original one. Such a growth model is therefore not self-dual like DP. The
quantities belonging to the dual process are designed in the following by a
hat (e.g.\ the mean particle density of the dual process is denoted by
$\hat{\rho}$).

As a last point we note the asymptotic scaling properties of the Green's
functions, the cumulants of the variables $n$ and $\tilde{n}$. One obtains
\begin{eqnarray}
\langle\prod_{i=1}^{N}n(\mathbf{x}_{i},t_{i})\prod_{j=1}^{\tilde{N}}\tilde
{n}(\mathbf{x}_{j},t_{j})\rangle^{(\mathrm{cum)}}  &  =\left\vert
\tau\right\vert ^{N\beta+\tilde{N}\tilde{\beta}}\nonumber\\
&  \hspace{-4cm}\times F_{N,\tilde{N}}(\{\left\vert \tau\right\vert ^{\nu
}\mathbf{x,}\left\vert \tau\right\vert ^{\nu z}t\},h/\left\vert \tau
\right\vert ^{(d+z)\nu-\tilde{\beta}},\rho_{0}/\left\vert \tau\right\vert
^{d\nu-\tilde{\beta}}) \label{GenSkal} 
\end{eqnarray}
with universal functions $F_{N,\tilde{N}}$. Particular scaling laws follow
from this asymptotic result of the renormalization group equation. It shows
that in general the scaling of growth processes is defined by four independent scaling exponents
$z,\nu,\beta,\tilde{\beta}$ \cite{MDHM94}. However, for theories which are self-dual like DP
we have in addition the relation $\beta=\tilde{\beta}$.

\section{Percolation and survival probabilities for directed percolation}

The expressions for the percolation and the survival probability for DP which
are constructed in this section, can also be used, \emph{mutatis mutandis},
for other growth processes, briefly considered in a later section.

Percolative spreading processes with absorbing states like DP provide two
independent fundamental order parameters, the density of the active particles
in the steady state $\rho_{st}$, and the percolation probability $P_{\infty}$.
Near the transition point they scale as $\rho_{st}\sim\left\vert
\tau\right\vert ^{\beta}$ and $P_{\infty}\sim\left\vert \tau\right\vert
^{\beta^{\prime}}$ where, in general, the two exponents $\beta$ and
$\beta^{\prime}$ are distinct from each other \cite{MDHM94}. It is well known that this
asymptotic scaling behavior follows from the relations $\rho_{st}\sim\langle
n\rangle$ and $P_{\infty}\sim\langle\tilde{n}\rangle$, where, in particular,
one has to specify the meaning of the expectation value for the latter
expectation value. Due to the duality transformation (\ref{DualTr}),
$P_{\infty}$ is intimately related to the particle density of the dual
process: $P_{\infty}\sim\hat{\rho}_{st}$, so that $\beta^{\prime}=\tilde
{\beta}=\hat{\beta}$. Hence, we get for self dual processes (like DP) the well
known relation $\beta^{\prime}=\beta$. In order to calculate scaling functions
and universal amplitude ratios it is, however, not sufficient to consider only
the asymptotic scaling behavior. We have to derive the exact relation between
the two order parameters.

The discussion of the cluster probabilities is complicated by the fact that
one has to deal carefully with the probabilities coming from the non-scaling
small clusters, say, with particle numbers smaller than $N_{0}$. For these
small clusters the asymptotic continuum field theory cannot be used. However,
the small clusters do not induce critical singularities due to their
finiteness. We assume as usual that they provide only analytical corrections
\cite{StAh94}.

Let $p_{N}(\tau)$ be the probability for a directed cluster in $(d+1)$
-dimensional space-time consisting of $N$ particles created from one particle,
the seed, at the origin of the space-time lattice, which is described by the
source (\ref{AnfQu}). For $N>N_{0}\gg1$ one can use the continuum
approximation
\begin{equation}
p_{N}(\tau)\approx P_{N}(\tau)=\langle\delta(N-\mathcal{N})\exp\bigl(\tilde
{n}(\mathbf{0},0)\bigr)\rangle\label{AsymProb} 
\end{equation}
where we have defined
\numparts
\begin{eqnarray}
\mathcal{N}  &  :=\int_{0}^{\infty}\rmd t\,\dot{\mathcal{N}}(t)\,,\label{N_fluk}\\
\dot{\mathcal{N}}(t)  &  :=\int \rmd^{d}x\,n(\mathbf{x},t)\,. \label{der-N} 
\end{eqnarray}
\endnumparts
The average in equation (\ref{AsymProb}) is performed with the response
functional, equation (\ref{DynFu_n}). One expects the asymptotic scaling behavior
$P_{N}(\tau)=N^{-x}f(\tau N^{y})$ \cite{StAh94}. In many cases,
e.g.\ absorptive noise as in DP near the critical point, and $N\gg1$, it is
possible to expand the exponential $\exp(\tilde{n}(\mathbf{0},0))$ in
equation (\ref{AsymProb}) in a power series to get asymptotic results.
Equation  (\ref{RandB}) leads to $\langle F[n]\rangle=F[0]$ for any functional $F$
of $n$. Hence, the first expansion term of the exponential, the $1$, inserted
into (\ref{AsymProb}) in place of $\exp\tilde{n}(\mathbf{0},0)$ leads exactly
to $\delta(N)=0$ if $N\gg1$. The higher powers of $\tilde{n}(\mathbf{0},0)$
contribute non-leading corrections to the asymptotic scaling behavior. Hence,
only the term $\tilde{n}(\mathbf{0},0)$ of the expansion must be retained.

We define a generating function
\begin{equation}
F(\tau,k)=\sum_{N=0}^{\infty}p_{N}(\tau)\rme^{-kN}\,,\quad k>0\,.
\label{GenFu} 
\end{equation}
This function has the property
\begin{equation}
F(\tau,k\rightarrow+0)=\left\{
\begin{array}
[c]{ccc} 
1 & \quad {\rm for}\quad & \tau\geq0\,,\\
1-P_{\infty} & \quad {\rm for}\quad & \tau<0\,.
\end{array}
\right.  \label{GenFu_lim} 
\end{equation}
From the continuous counterpart of (\ref{GenFu}) we obtain, neglecting
possible analytic corrections that arise from the small clusters
\begin{eqnarray}
F  &  (\tau,k)-F(\left\vert \tau\right\vert ,k)=\int_{0}^{\infty
}\rmd N\,\bigl(P_{N}(\tau)-P_{N}(\left\vert \tau\right\vert )\bigr)\rme^{-kN}\nonumber\\
&  =\langle\exp\bigl(\tilde{n}(\mathbf{0},0)-k\mathcal{N}\bigr)\rangle
(\tau)-\langle\exp\bigl(\tilde{n}(\mathbf{0},0)-k\mathcal{N}\bigr)\rangle
(\left\vert \tau\right\vert )\,\,. \label{Eff} 
\end{eqnarray}
Here we rely on the property that divergencies like $P_{N}\sim N^{-x}$ for
small $N$ are independent of $\tau$ and cancelled in the integrand of
equation (\ref{Eff}) \cite{StAh94}. Using now the fact that in the subcritical
absorbing phase with unbroken symmetries $\lim_{k\rightarrow+0}\langle
\rme^{-k\mathcal{N}}\exp\tilde{n}(\mathbf{0},0)\rangle(\tau
>0)=\langle\exp\tilde{n}(\mathbf{0},0)\rangle(\tau>0)=1$, we arrive at
\begin{equation}
P_{\infty}=1-\lim_{k\rightarrow+0}\langle\exp\bigl(\tilde{n}(\mathbf{0}
,0)-k\mathcal{N}\bigr)\rangle\,. \label{P-inf} 
\end{equation}
Provided that the asymptotic expansion of the exponential is allowed we get
\begin{equation}
P_{\infty}=-\lim_{k\rightarrow+0}\langle\rme^{-k\mathcal{N}}\tilde
{n}(\mathbf{0},0)\rangle\propto\left\vert \tau\right\vert ^{\tilde{\beta}}
\label{LimP} 
\end{equation}
by applying equation (\ref{GenSkal}). With the help of the duality transformation
(\ref{DualTr}) in conjunction with the time translation invariance, we obtain
finally
\begin{eqnarray}
P_{\infty}  =\alpha^{2}\lim_{h\rightarrow+0}\langle n(\mathbf{0}
,\infty)\rme^{\lambda h\widetilde{\mathcal{N}}}\rangle^{(\mathrm{dual}
)}=\alpha^{2}\hat{\rho}_{st}\propto\left\vert \tau\right\vert ^{\hat{\beta}}
\label{RelPr} 
\end{eqnarray}
where $\widetilde{\mathcal{N}}=\int \rmd t\int \rmd^{d}x\,\tilde{n}(\mathbf{x},t)$ and
$\lambda h=k/\alpha^{2}$ constitutes a constant particle source that is taken
to zero at the end of the calculation. Note that the particle density of the
dual process $\hat{\rho}_{st}$ is independent of any initial state unequal to
the vacuum. In self-dual theories like DP, $\hat{\beta}
=\beta$ and the parameter $\alpha$ can be chosen that $\hat{\rho}_{st}=\rho_{st}$, as shown above. Then, we see that the redundant variable $\alpha$ enters the relation
between the two order parameters: $P_{\infty}=\alpha^{2}{\rho}_{st}$

Now we turn our attention to time dependent observables. As above, there are
two fundamental complementary quantities, namely the particle density
$\rho(t)$ starting from a fully occupied initial state at time $t=0$, and the
survival probability $P(t)$ that a cluster grown from a single seed is still
active at time $t$. Both are monotonically decreasing functions of time. The
exact relation between the two is found with the help of the following consideration.

Let $p_{N}(t)$ denote the probability that a cluster has exactly $N$ active
particles at time $t$. We define the function
\begin{equation}
P(k,t)=\sum_{N=0}^{\infty}\bigl(1-\rme^{-kN}\bigr)p_{N}(t)\,.
\label{GenFu_t} 
\end{equation}
Of course, we have $\lim_{k\rightarrow\infty}P(k,t)=P(t)$. We distinguish
again between small and large clusters. The former lead to uncritical
corrections. For the large clusters we use the approximation
\begin{equation}
p_{N}(t)\approx P_{N}(t)=\langle\delta(N-\dot{\mathcal{N}}(t))\exp\tilde
{n}(\mathbf{0},0)\rangle\label{AsymProb_t} 
\end{equation}
where $\dot{\mathcal{N}}(t)$ is defined in equation (\ref{der-N}). Thus, we get
from the continuous counterpart of equation (\ref{GenFu_t})
\begin{eqnarray}
\fl \quad P(k,t)=\int_{0}^{\infty}\rmd N\,\bigl(1-\rme^{-kN}\bigr)P_{N}(t) =\langle\exp\tilde{n}(\mathbf{0},0)\rangle-\langle\exp\bigl(\tilde
{n}(\mathbf{0},0)-k\dot{\mathcal{N}}(t)\bigr)\rangle\,. \label{Lapl_P} 
\end{eqnarray}
The first term of the last equation, $\langle\exp\tilde{n}(\mathbf{0}
,0)\rangle$, is equal to one. Hence, we obtain the survival probability
\begin{equation}
P(t)=1-\lim_{k\rightarrow\infty}\langle\exp\bigl(\tilde{n}(\mathbf{0}
,0)-k\dot{\mathcal{N}}(t)\bigr)\rangle\,. \label{SProb} 
\end{equation}
After expanding the exponential to get asymptotic results, and utilizing the
time translation invariance of the process we arrive at
\begin{equation}
P(t)=-\lim_{k\rightarrow\infty}\langle\tilde{n}(\mathbf{0},-t)\rme
^{-k\dot{\mathcal{N}}(0)}\rangle\propto t^{-\tilde{\beta}/\nu z}\,.
\label{LimP_t} 
\end{equation}
Using again the duality transformation (\ref{DualTr}), we find
\begin{eqnarray}
P(t) =\alpha^{2}\lim_{\hat{\rho}_{0}\rightarrow\infty}\langle
n(\mathbf{0},t)\rme^{\hat{\rho}_{0}\dot{\widetilde{\mathcal{N}}}
(0)}\rangle^{(\mathrm{dual})}\nonumber =\alpha^{2}\hat{\rho}(t)\propto t^{-\hat{\beta}/\hat{\nu}\hat{z}}\,,
\label{RelPr_t} 
\end{eqnarray}
where $\dot{\widetilde{\mathcal{N}}}(0)=\int \rmd^{d}x\,\tilde{n}(\mathbf{x},0)$,
and $\hat{\rho}_{0}=k/\alpha^{2}$ constitutes an initial constant particle
density. Hence, $\hat{\rho}(t)$ is the particle density of the dual process
at time $t$ starting from a infinite homogeneous initial density $\hat{\rho
}_{0}$, corresponding to a fully occupied initial state. For self-dual
processes like DP we have in fact $\hat{\rho}(t)=\rho(t)$ and $\hat{\beta
}/\hat{\nu}\hat{z}=\beta/\nu z$. Relation (\ref{RelPr_t}) is fully analogous
to the steady state relation (\ref{RelPr}), but note the different limits
$h\rightarrow0$ and $\hat{\rho}_{0}\rightarrow\infty$. In any case it follows
from $\lim_{t\rightarrow\infty}\hat{\rho}(t)=\hat{\rho}_{st}$ that
\begin{equation}
\lim_{t\rightarrow\infty}P(t)=P_{\infty}\,. \label{Limites} 
\end{equation}

The expressions (\ref{P-inf}, \ref{LimP}, \ref{SProb}, \ref{LimP_t}) and the
relations (\ref{RelPr}, \ref{RelPr_t}) are the main general results of this
paper. They hold near criticality and for not too small times $t$, because we
have neglected several analytic corrections and used, for discrete processes,
the continuum approximation. Also we have neglected in (\ref{LimP},
\ref{LimP_t}) higher powers of $\tilde{n}$. Being interested in non-asymptotic
results one has to use the full expression $(\exp\tilde{n}-1)$ instead of the
single term $\tilde{n}$ (see (\ref{P-inf}) and (\ref{SProb})).

\section{Probabilities in a mean field approximation}

To demonstrate the applicability of our results (\ref{LimP}, \ref{LimP_t}), we
calculate the corresponding path-integrals in a mean-field approximation. Then
we are able to write down explicit simple expressions for all quantities and
relationships. At first we determine the asymptotic expression of the cluster
probability (\ref{AsymProb}). Instead of directly evaluating this quantity we
consider the Laplace transform
\begin{eqnarray}
\tilde{M}  &  (k,\tau)=\int_{0}^{\infty}\rmd N\,\rme^{-kN}P_{N}
(\tau)=\langle\exp\bigl(\tilde{n}(\mathbf{0},0)-k\mathcal{N}\bigr)\rangle
\nonumber\\
&  =\int\mathcal{D}(\tilde{n},n)\mathcal{\,}\exp\bigl(-\mathcal{J}[\tilde
{n},n]-(k,n)+\tilde{n}(\mathbf{0},0)\bigr)\,. \label{M-tilde} 
\end{eqnarray}
The saddle point of the exponential weight follows from the solutions of the
variational equations
\numparts
\begin{eqnarray} \label{SPEq} 
\fl \qquad\qquad \frac{\delta\bigl(\mathcal{J}[\tilde{n},n]+(k,n)\bigr)}{\delta n(\mathbf{x}
,t)} =-\frac{\partial\tilde{n}}{\partial t}+\lambda(\tau-\nabla^{2}
)\tilde{n}+\lambda g^{\prime}n\tilde{n}-\frac{\lambda g^{\prime\prime}}{2}\tilde{n}^{2}+k=0\,,\label{SPEq-1}\\
\fl \qquad\qquad  \frac{\delta\bigl(\mathcal{J}[\tilde{n},n]+(k,n)\bigr)}{\delta\tilde
{n}(\mathbf{x},t)} =\frac{\partial n}{\partial t}+\lambda(\tau-\nabla
^{2})n+\frac{\lambda g^{\prime}}{2}n^{2}-\lambda g^{\prime\prime}n\tilde{n}=0\,. \label{SPEq-2} 
\end{eqnarray}
\endnumparts
Replacing $k$ by $k\,\theta(T-t)$ where the limit $T\rightarrow\infty$ is
performed at the end, the solutions of the saddle point equations, subject to
the constraints (\ref{RandB}), are
\numparts
\begin{eqnarray} \label{tSpSol} 
\tilde{n}(\mathbf{x},t)|_{s.p.}  &  =\frac{-2\theta(T-t)\bigl(\exp(\lambda
w(T-t))-1\bigr)k}{\lambda(w+\tau)\exp(\lambda w(T-t))+(w-\tau)}
\,,\label{tSPSol-1}\\
n(\mathbf{x},t)|_{s.p.}  &  =0 \,, \label{tSpSol-2} 
\end{eqnarray}
\endnumparts
where $w=\sqrt{\tau^{2}+2g^{\prime\prime}k/\lambda}$. Consequently we have
$(\mathcal{J}[\tilde{n},n]+(k,n))|_{s.p.}=0$, and we obtain in the limit
$T\rightarrow\infty$
\begin{eqnarray}
\tilde{M}(k,\tau) =\exp\tilde{n}(\mathbf{0},0)|_{s.p.} =\exp\Big(\bigl(\tau-\sqrt{\tau^{2}+2g^{\prime\prime}k/\lambda
}\bigr)/g^{\prime\prime}\Big)\,. \label{M(k,t)} 
\end{eqnarray}
Using equation (\ref{P-inf}), the limit $k\rightarrow+0$ of this expression yields
the percolation probability
\begin{eqnarray}
\fl \qquad\qquad P_{\infty}(\tau) =1-\lim_{k\rightarrow+0}\tilde{M}(k,\tau)=1-\exp\bigl((\tau-\left\vert \tau\right\vert )/g^{\prime\prime
}\bigr)\approx\frac{2\left\vert \tau\right\vert }{g^{\prime\prime}}\,\theta
(-\tau)\,. \label{mfPerk} 
\end{eqnarray}
The last expression is in accordance with equation (\ref{RelPr}), since $\rho_{st}(\tau)=2\left\vert \tau\right\vert\,\theta(-\tau)/g^{\prime}$ is the stationary solution of the mean field equation.

From equation (\ref{M-tilde}) we find the cluster probability $P_{N}$ by an
inverse Laplace transform of $\tilde{M}(k,\tau)$
\begin{eqnarray}
\fl \qquad P_{N}(\tau)=\frac{1}{2\pi i}\int_{\sigma-i\infty}^{\sigma+i\infty
}\rmd k\,\rme^{kN}\tilde{M}(k,\tau)=\frac{\rme^{\tau/g^{\prime\prime}}N^{-3/2}}{\sqrt{2\pi\lambda
g^{\prime\prime}}}\,\exp\Big(-\frac{1}{2g^{\prime\prime}}\bigl(\lambda\tau
^{2}N+(\lambda N)^{-1}\bigr)\Big)\nonumber\\
\approx\frac{N^{-3/2}}{\sqrt{2\pi\lambda g^{\prime\prime}}}\,\exp
\bigl(\tau/g^{\prime\prime}-\lambda\tau^{2}N/2g^{\prime\prime}
\bigr) \label{P(N)} 
\end{eqnarray}
The last row presents indeed the well known asymptotic probability
distribution for the generation of a directed $N$-cluster on the Bethe-lattice
\cite{StAh94}. $P_{N}(\tau)$ shows the symmetry property
\begin{equation}
P_{N}(-\left\vert \tau\right\vert )=\bigl(1-P_{\infty}(-\left\vert
\tau\right\vert )\bigr)\,P_{N}(\left\vert \tau\right\vert ) \label{ab-bel} 
\end{equation}
between the probability distributions above and below the percolation point with $\int
_{0}^{\infty}\rmd N\,P_{N}(\left\vert \tau\right\vert )=1$.

Now we turn our attention to the survival probability (\ref{SProb}). We have
to solve the saddle-point equations (\ref{SPEq}) where $k$ is replaced by
$k\,\delta(t)$ or, in other words, with the final-condition $\tilde
{n}(\mathbf{x},0)=-k$ and $k=0$ for $t\neq0$. We find
\numparts
\begin{eqnarray} \label{SPSol2}
\tilde{n}(\mathbf{x},-t)|_{s.p.}  &  =\frac{-2\tau\theta(t)\exp(-\lambda\tau
t)}{2\tau k^{-1}+g^{\prime\prime}\bigl(1-\exp(-\lambda\tau t)\bigr)}
\,,\label{SPSol2-1}\\
n(\mathbf{x},-t)|_{s.p.}  &  =0\,. \label{SPSol2-2} 
\end{eqnarray}
\endnumparts
Hence, according to equation (\ref{SProb}), the survival probability in mean field
approximation is given by
\begin{eqnarray}
\fl \qquad P(t)=1-\exp\Big(\frac{-2\tau/g^{\prime\prime}}{\exp(\lambda\tau
t)-1}\Big)\approx\frac{2\tau/g^{\prime\prime}}{\exp(\lambda\tau t)-1}
\nonumber\\
\qquad \hspace{-0.5cm}=\left\{
\begin{array}
[c]{lll} 
2/(\lambda g^{\prime\prime}t) & \ {\rm for\ } & \tau=0,\\
(2\tau/g^{\prime\prime})\exp(-\lambda\tau t) & \ {\rm for\ } & \tau
>0,\ t\rightarrow\infty\,,\\
(2\left\vert \tau\right\vert /g^{\prime\prime})\bigl(1+\exp(-\lambda\left\vert
\tau\right\vert t)\bigr) & \ {\rm for\ } & \tau<0,\ t\rightarrow\infty\,.
\end{array}
\right.  \label{P(t)} 
\end{eqnarray}
We see that $P(t)$ owns all the required properties displayed in
equations (\ref{RelPr_t}, \ref{Limites}).

According to equation (\ref{Lapl_P}) the inverse Laplace transform of
\begin{eqnarray}
1-P(k,t)  =\langle\exp\bigl(\tilde{n}(\mathbf{0},-t)-k\dot{\mathcal{N}
}(0)\bigr)\rangle=\exp\tilde{n}(\mathbf{0},-t)|_{s.p.} 
\end{eqnarray}
leads back to
\begin{eqnarray}
\fl \qquad P_{N}(t)=\langle\delta(N-\dot{\mathcal{N}}(t))\exp\tilde{n}
(\mathbf{0},0)\rangle=\frac{1}{2\pi i}\int_{\sigma-i\infty}^{\sigma+i\infty}
\rmd k\,\bigl(1-P(k,t)\bigr)\rme^{kN}\nonumber\\
=P_{0}(t)\delta(N)+\frac{1}{2\pi i}\int_{\sigma-i\infty}^{\sigma+i\infty
}\rmd k\,\bigl(P(\infty,t)-P(k,t)\bigr)\rme^{kN}\nonumber\\
=:P_{0}(t)\delta(N)+\bar{P}_{N}(t)\,. \label{P(N,t)1} 
\end{eqnarray}
$P_{0}(t)=1-P(t)$ is the extinction probability, and we have used
equation (\ref{SProb}). The subtraction in the integral that defines $\bar{P}
_{N}(t)$ is needed for the application of Jordan's lemma. Only after this
subtraction the inverse Laplace transform can be calculated by closing the
integration path with a large semi-circle which tends to infinity in the left
complex $k$-plane. By this subtraction-procedure the $\delta$-contribution of
the absorbing state is splitted off. Using the saddle point solution
(\ref{SPSol2-1}) we find
\numparts
\begin{eqnarray} \label{P-Nmf}
\fl \qquad P_{0}(t)  &  =\exp\bigl(-A(t)\bar{N}(t)\bigr)\,,\label{P-Nmf-1}\\
\fl \qquad \bar{P}_{N}(t)  &  =A(t)\sqrt{\bar{N}(t)/N}\,I_{1}\Bigl(2A(t)\sqrt{\bar
{N}(t)N}\Bigr)\exp\bigl[-A(t)\bigl(\bar{N}(t)+N\bigr)\bigr]\,. \label{P-Nmf-2} 
\end{eqnarray}
\endnumparts
Here we have defined
\numparts
\begin{eqnarray}
A(t)  &  =\frac{2\tau/g^{\prime\prime}}{1-\exp(-\lambda\tau t)}\,,\\
\bar{N}(t)  &  =\exp(-\lambda\tau t)\,.
\end{eqnarray}
\endnumparts
Note that with the identifications $\lambda=1$ and $g^{\prime\prime}=2$ the
mean field solution (\ref{P-Nmf}) turns into the distribution calculated with
the Fokker-Planck equation (\ref{abFPSol}) with $x_{0}=1$ as we have
anticipated in section II.

Asymptotically, if $A(t)\bar{N}(t)\ll1$ and $A(t)N$ is finite, we can expand
equations (\ref{P-Nmf}) and obtain
\begin{equation}
P_{N}(t)\approx\bigl(1-A(t)\bar{N}(t)\bigr)\delta(N)+A(t)^{2}\bar{N} 
(t)\,\exp\bigl(-A(t)N\bigr)\,. \label{P(N)_as} 
\end{equation}
This expression can also be derived if $\exp\tilde{n}(\mathbf{0},0)$ is
approximated by $1+\tilde{n}(\mathbf{0},0)$ in equation (\ref{P(N,t)1}).

For logarithmic corrections to the mean field results in the upper critical
dimension $d_{c}=4$ see \cite{JaSt04}.

\section{Multiplicative noise and comparison with other work}

Now we equip the Langevin equation (\ref{LangGl}) with multiplicative noise
\begin{equation}
\overline{\zeta(\mathbf{x},t)\zeta(\mathbf{x}^{\prime},t^{\prime})}=\lambda
g^{\prime\prime}n(\mathbf{x},t)^{2}\delta(\mathbf{x}-\mathbf{x}^{\prime
})\delta(t-t^{\prime}) \label{multNoise} 
\end{equation}
instead of the absorptive one, equation (\ref{LangQu}). It is well known that in this
case the Langevin equation (\ref{LangGl}) cannot represent a physical
reaction-diffusion system \cite{HoTa97,Mu98}. At least diffusional noise must
be included and the response functional changes to
\begin{eqnarray}
\fl \qquad \mathcal{J}^{\prime}[\tilde{n},n] =\int \rmd t\,\rmd^{d}x\,\biggl\{\lambda
\tilde{n}\Big(\lambda^{-1}\frac{\partial}{\partial t}+(\tau-\nabla
^{2})+\frac{g^{\prime}}{2}n-\frac{g^{\prime\prime}}{2}\tilde{n}n\Big)n-\lambda
n(\nabla\tilde{n})\biggr\}\,. \label{J_m} 
\end{eqnarray}
The upper critical dimension $d_{c}$ decreases from $4$ to $2$. The naive
scaling dimensions of the fields become $d$ for $n$ and $0$ for $\tilde{n}$.
Hence, one expects that also non-Gaussian noise contributions becomes
relevant. Indeed the field theory based on $\mathcal{J}^{\prime}$, equation (\ref{J_m}),
is not renormalizable as it stands. However, a quasicanonical transformation
to \textquotedblleft bosonic\textquotedblright\ fields
\numparts
\begin{eqnarray} \label{Boson}
a^{+}  &  =1+\tilde{a}=\exp\tilde{n}\,,\label{Erz}\\
a  &  =n\exp(-\tilde{n})\,, \label{Vern} 
\end{eqnarray}
\endnumparts
leads to the elimination of the diffusional noise term. The resulting action
is known from the theory of branching and annihilating random walks
\cite{CaTa96}. The corresponding Langevin equation is formally equipped with
imaginary multiplicative noise.

It is instructive to apply the saddle point approximation to the response
functional (\ref{J_m}). This procedure is expected to be correct above the
upper critical dimension. The appropriate saddle point equations are given by
(\ref{SPEq}) with the replacement of $g^{\prime\prime}$ by $g^{\prime\prime} 
n$, and an additional contribution arising from the diffusional noise.
However, both terms are cancelled for homogeneous solutions with $n=0$. We get
the time-dependent solution
\begin{equation}
\tilde{n}(\mathbf{x},-t)|_{s.p.}=-k\,\exp(-\lambda\tau t)\,. \label{SPSol-m} 
\end{equation}
which goes to infinity for $k\rightarrow\infty$. Thus, for each finite time
$t$ we obtain from equation (\ref{SProb})
\begin{equation}
P(t)=1-\lim_{k\rightarrow\infty}\exp\tilde{n}(\mathbf{x},-t)|_{s.p.} =1.
\label{LimP_t-m} 
\end{equation}
Hence, the survival probability for processes equipped with multiplicative
behaves trivially at least in the saddle point approximation. It is obvious
that an expansion of $\exp\tilde{n}$ is not allowed. Indeed, it is appropriate
to change to the bosonic variables (\ref{Boson}).

Now we are in a position to compare our result for the survival probability
with the corresponding expression given by MGT. As we have seen,
the distribution function $P_{N}(t)$ (\ref{P(N,t)1}) consists of two parts
\begin{eqnarray}
P_{N}(t)  &  =P_{N}(t)=\langle\delta(N-\dot{\mathcal{N}}(t))\exp\tilde
{n}(\mathbf{0},0)\rangle\nonumber\\
&  =\bigl(1-P(t)\bigr)\delta(N)+P(t)\bar{p}_{N}(t) \label{P(N,t)} 
\end{eqnarray}
where the nonsingular distribution $\bar{p}_{N}(t)$ is normalized to $\int
_{0}^{\infty}\rmd N\,\bar{p}_{N}(t)=1$. The survival probability $P(t)$ is a
monotonic decreasing function in the case of absorptive noise and equal to one
in the case of multiplicative one. MGT have introduced the expression
\begin{equation}
P^{(MGT)}(t,\alpha)=\langle\theta(\dot{\mathcal{N}}(t)-\alpha)\tilde
{n}(\mathbf{0},0)\rangle\ \label{MGT} 
\end{equation}
as a substitute of the survival probability. In the case of absorptive noise,
where asymptotically the expansion of $\exp\tilde{n}$ is allowed, we get from
equation (\ref{P(N,t)})
\begin{equation}
\langle\delta(N-\dot{\mathcal{N}}(t))\tilde{n}(\mathbf{0},0)\rangle
=P(t)\bigl[\bar{p}_{N}(t)-\delta(N)\bigr]\,.
\end{equation}
Hence, one sees immediately
\begin{equation}
\lim_{\alpha\rightarrow0}P^{(MGT)}(t,\alpha)=P(t)\,
\end{equation}
in the large $t$ and small $\tau$ limit. The advantage of our expression
(\ref{LimP_t}) for the survival probability is its calculability in
perturbation theory. This contrasts to the substitute (\ref{MGT}), in which
the Heaviside-$\theta$-function is a insurmountable complication for any calculation.

Things become worse for multiplicative noise. As we have shown, the expansion of $\exp\tilde{n}$
is dubious, and the survival probability $P(t)=1$ as long as $t$ is finite.
Even if the distribution $\bar{p}_{N}(t)$ piles up near $N=0$ with $\bar
{p}_{N}(t)\sim N^{\rho}$ for small $N$ and a width $B(t)$, the substitute
behaves like
\begin{equation}
P^{(MGT)}(t,\alpha)=1+O\bigl((\alpha/B(t))^{\rho+1}\bigr)
\end{equation}
and goes to zero exponentially if $\alpha\gg B(t)$. Hence, $\alpha$ is a
dangerous variable, and it is not clear which value should be assigned to it.

\section{Other growth models with absorptive noise}

In this section we will briefly discuss the applicability of our expressions
to other growth processes with absorbing states widely discussed in the
literature. We present several variants of the original DP, for which the
expressions (\ref{LimP},\ref{LimP_t}) for the probabilities are well applicable.

\subsection{Tricritical directed percolation}

Tricritical directed percolation (TDP) can be understood as the limit of DP in
which the renormalized coupling constant $g^{\prime}$ goes to zero. In this
limit no growth limiting term shows up in the original mean field part of the
equation of motion of DP (\ref{LangGl},\ref{R_DP}). We expect that formerly
irrelevant couplings become now relevant. The nature of these terms depends on
the physics of the growth process under consideration. The simplest case is
the occurrence of a higher order effective coupling in the reaction rate. In
TDP \cite{OhKe87,Ja87} one assumes the form
\begin{equation}
R\left[  n\right]  =\lambda\Big(\tau+\frac{g^{\prime}}{2}n+\frac{f}{6} 
n^{2}\Big) \label{R_TDP} 
\end{equation}
that leads to an interaction part of the response functional (\ref{DynFu_n})
\begin{equation}
\mathcal{V}_{TDP}[\tilde{n},n]=\frac{1}{2}\Big(g^{\prime}n+\frac{f}{3} 
n^{2}-g^{\prime\prime}\tilde{n}\Big). \label{V_TDP} 
\end{equation}
In the mean field approximation, the tricritical point which separates the
continuous DP-transition from a discontinuous one is found at $\tau=g^{\prime
}=0$. The noise term has the absorbing form (\ref{LangQu}). However, the dual
symmetry (\ref{DualTr}) is lost and we expect different exponents $\beta
\neq\tilde{\beta}=\beta^{\prime}$. Indeed, a field-theoretic renormalization
group calculation up to 2-loop order in an $\varepsilon$-expansion below the
upper critical dimension $d_{c}=3$, $\varepsilon=3-d$, yields
\begin{equation}
\beta=\frac{1}{2}-0.458\varepsilon+O(\varepsilon^{2}),\quad\beta^{\prime
}=1+O(\varepsilon^{2}). \label{b_TDP} 
\end{equation}
The other exponents are $z=2+0.0086\,\varepsilon+O(\varepsilon^{2})$ and
$\nu=1/2+0.0075\,\varepsilon+O(\varepsilon^{2})$. The crossover exponent
$\phi=1/2-0.012\,\varepsilon+O(\varepsilon^{2})$ describes the scaling of the
crossover from TDP to DP. It can be easily seen that the mean-field
calculations of the probabilities in section V lead to the same results for
TDP as found for DP.

\subsection{Compact directed percolation}

The universality class of compact directed percolation (CDP) that includes
e.g.\ the voter model \cite{Li85,DoKi84,Es89,DiYu95}, describes another
universality class that can be reached in the limit $g^{\prime}\rightarrow0$
of DP. In contrast to TDP, in CDP a limiting term in the deterministic part of
the continuum Langevin equation of motion (\ref{LangGl},\ref{R_DP}) does not
exist a priori. Thus, in the active phase with $\tau<0$, the density grows to
infinity corresponding to a state with a completely filled lattice.
Renormalizing the density of the filled lattice to the value $n=1$, we expect
that formerly irrelevant couplings even in the noise become relevant. A
special physical condition comes from the demand that the empty and the filled
lattice are both absorbing states and the exchange $n\rightarrow1-n$,
$\tau\rightarrow-\tau$ is a symmetry. The patchy spatial structure of the
CDP-states motivates an expansion of the equation of motion in the small
quantity $n(1-n)$. Hence, the relevant terms of the reaction rate of the
Langevin equation (\ref{LangGl},\ref{R_DP}) are given by
\begin{equation}
R\left[  n\right]  _{(\mathbf{x},t)}n(\mathbf{x},t)=\lambda\tau
\bigl(1-n(\mathbf{x},t)\bigr)n(\mathbf{x},t). \label{R_VM} 
\end{equation}
The reaction rate is zero at the critical point $\tau=0$. For $\tau\neq0$, the
model has a bias which prefers the state $n=0$ or $n=1$ depending on the sign
of $\tau$. The noise in the empty or the filled lattice is zero. This enforces
Gaussian fluctuations of the noise of the form \cite{DiYu95}
\begin{equation}
\overline{\zeta(\mathbf{x},t)\zeta(\mathbf{x}^{\prime},t^{\prime})}=\lambda
gn(\mathbf{x},t)\bigl(1-n(\mathbf{x},t)\bigr)\delta(\mathbf{x}-\mathbf{x}
^{\prime})\delta(t-t^{\prime}). \label{LangQ_VM} 
\end{equation}
equations (\ref{R_VM},\ref{LangQ_VM}) lead to the response functional
(\ref{DynFu_n}) of CDP
\begin{eqnarray}
\fl \qquad \mathcal{J}_{CDP}[\tilde{n},n] =\int \rmd t\,d^{d}x\,\biggl\{\lambda\tilde
{n}\Big(\lambda^{-1}\frac{\partial n}{\partial t}-\nabla^{2}n+\tau n(1-n)-\frac{g}{2}\tilde{n}n(1-n)\Big)\biggr\} \label{J_VM} 
\end{eqnarray}
This response functional is invariant under the transformation $n\rightarrow
1-n$, $\tilde{n}\rightarrow-\tilde{n}$, $\tau\rightarrow-\tau$. For $\tau=0$,
$\mathcal{J}_{CDP}[\tilde{n},n]$ transforms under duality (\ref{DualTr}) to an
action which describes annihilating random walks \cite{Pe86,Le94}. Note that
the noise then becomes imaginary in the corresponding Langevin equation.
Physically this results from the anticorrelations present in the annihilating
walk process. Consequently, the notion of density variables are inappropriate
in such a case. Instead the fields $n,\tilde{n}$ represent annihilation and
creation operators of a bosonic field theory. The scaling dimensions of
$\tilde{n}$ and $n$ are not related, and are given trivially by the naive
dimensions which are $d$ and $0$, respectively. The upper critical dimension
is $d_{c}=2$. The renormalization of the bias $\tau$ is intimately related to
the exactly computable renormalization of the coupling constant $g$. A short
presentation of the field theory is given in the appendix.

We note here the well known order parameter exponents of CDP for $d\leq2$
\begin{equation}
\beta=0,\quad\beta^{\prime}=1. \label{Exp_VM} 
\end{equation}
The dynamical and the correlation length exponents are $z=2$ and $\nu=1/d$.
The crossover exponent to DP is found to be $\phi=2/d$.

For $d\geq2$, the mean-field results for the probabilities derived for DP are
applicable. The survival probability for clusters generated from a localized
seed in an otherwise empty lattice at and below $d=2$, as well as the
percolation probability follows from our general formulas (\ref{LimP}) and (\ref{LimP_t}) to
\begin{equation}
P(t)=t^{-d/2}f(\tau t^{d/2}),\quad P_{\infty}\propto\left\vert \tau\right\vert
. \label{Prob_VM} 
\end{equation}

\subsection{Conserved diffusing secondary field}

In analogy to the extension of Model A to Model C from the Halperin-Hohenberg
catalog of fundamental field-theoretic models for critical dynamics, one can
consider an extension of DP to DP-C \cite{Ja01b}. In Model C a conserved
diffusing density which is not critical itself is coupled to the relaxing
dynamics of an order parameter. In the DP-C universality class, a conserved
density $c(\mathbf{x},t)$ is coupled to the growing density field
$n(\mathbf{x},t)$ of the DP-process. The relevant Langevin dynamics is given
by
\numparts 
\begin{eqnarray} \label{DP-C}
\partial_{t}n  &  =\lambda\nabla^{2}n-\lambda\Big(\tau+\frac{g^{\prime}}
{2}n+fc\Big)n+\zeta,\\
\partial_{t}c  &  =\nabla^{2}\bigl(\gamma c+\sigma n\bigr)+\zeta^{\prime} 
\end{eqnarray}
\endnumparts
where the Langevin forces are Gaussian absorptive noise (\ref{LangQu}) and
Gaussian diffusion noise
\begin{equation}
\overline{\zeta^{\prime}(\mathbf{x},t)\zeta^{\prime}(\mathbf{x}^{\prime
},t^{\prime})}=-\gamma\nabla^{2}\delta(\mathbf{x}-\mathbf{x}^{\prime}
)\delta(t-t^{\prime}). \label{DP-C-noise} 
\end{equation}
The first model of this type (with $\sigma=0$) was introduced by Kree et
al.\ \cite{KSS89} to deal with a catalytic poison which may damage a growing
population. Ten years later, van Wijland et al.\ \cite{WOH00} have shown that
a special reaction-diffusion system ($A+B\rightarrow2B$, $B\rightarrow A$)
corresponds in general to DP-C with a cross-diffusion coefficient $\sigma
\neq0$. However, if $\sigma f>0$, the renormalization flow leads finally to a
violation of the stability bound $2\sigma f<g$ of the mean field part of
equations (\ref{DP-C}), which is interpreted as the occurrence of a fluctuation
induced first order transition. In this sense, the KSS model ($\sigma=0$)
defines a tricritical point of the critical DP-C model with $\sigma f<0$ and
with its own universality class. Note that the limit $\gamma\rightarrow0$
leads to another universality class \cite{Ro00} called the conserved lattice
gas (CLG) or Manna class \cite{Ma91}. The systems belonging to this
universality class exhibit infinitely many absorbing states characterized by
the different distributions of a conserved quantity. A recent renormalization
group approach \cite{Wi02} is dubious. It proposes an upper critical dimension
$d_{c}=6$, whereas simulations \cite{Lu01,LuHe03} clearly exhibit an upper
critical dimension $d_{c}=4$.

As the result of special symmetries, the DP-C model has the exact exponents
$z=2$ and $\nu=2/d$ below the upper critical dimension $d_{c}=4$. The special
KSS model is invariant under the duality transformation (\ref{DualTr})
extended by $c(\mathbf{x},t)\rightarrow c(\mathbf{x},-t)$. Hence, for
$\sigma=0$ we have the equality
\begin{equation}
\beta=\beta^{\prime}=1-\frac{\varepsilon}{16}+O(\varepsilon^{2}) \label{b-KSS} 
\end{equation}
where the expansion results from a 1-loop calculation\ \cite{KSS89}. In
general, for $\sigma\neq0$, dual symmetry is lost. In this case it turns out,
that $n$ does not renormalize. Its anomalous dimension is zero whereas the
anomalous dimension of $\tilde{n}$ is nontrivial. One obtains \cite{WOH00}
finally
\begin{equation}
\beta=1,\quad\beta^{\prime}=1-0.1065\,\varepsilon+O(\varepsilon^{2}).
\label{b-WOH} 
\end{equation}
Also for the DP-C as well as the CLG class, the mean field calculation of the
survival and percolation probability lead to the formulas known from DP and is
expected to be correct above the upper critical dimension.

\subsection{Dynamic isotropic percolation}

Contrary to the growth models considered up to now, the dynamic isotropic
percolation (dIP) modelled by the general epidemic process (GEP) \cite{Gr83}
leads to memory terms in the equation of motion for the growing density. The
debris (or the immunes in the language of the GEP) is generated by the agent,
i.e. the active particles (or the infected individuals). It suppresses the
growth process itself and leads finally to the extinction of the disease
($n=0$ in each finite spatial region). The spatial distribution of the debris
in the stationary limit is described by the usual static percolation
statistics. Thus, it is more appropriate to consider the density of the debris
as an order parameter field. It is proportional to the time integral of the
density of the active particles
\begin{equation}
m(\mathbf{x},t)=\lambda\int_{-\infty}^{t}\rmd t^{\prime}\,n(\mathbf{x},t^{\prime
}). \label{OP_IP} 
\end{equation}

The relevant terms in the reaction rate of the Langevin equation
(\ref{LangGl}) are now given by
\begin{equation}
R\left[  n\right]  _{(\mathbf{x},t)}=\lambda\bigl(\tau+g^{\prime} 
m(\mathbf{x},t)\bigr). \label{R_IP} 
\end{equation}
Therefore, the response functional is follows as \cite{Ja85}
\begin{eqnarray}
\fl \qquad\qquad \mathcal{J}_{dIP}[\tilde{n},n] =\int \rmd t\,\rmd^{d}x\,\biggl\{\lambda\tilde
{n}\Big(\lambda^{-1}\frac{\partial}{\partial t}+(\tau-\nabla^{2})+g^{\prime}m-\frac{g^{\prime\prime}}{2}\tilde{n}\Big)n\biggr\}.
\label{J_DIP} 
\end{eqnarray}
dIP is not invariant under the transformation (\ref{DualTr}). However, the
variant of the duality transformation
\begin{equation}
\alpha^{-1}\tilde{n}(\mathbf{x},t)\rightarrow-\alpha m(\mathbf{x},-t)
\label{DualTr_2} 
\end{equation}
with $\alpha=\sqrt{g^{\prime}/g^{\prime\prime}}$ transforms the response
functional (\ref{J_DIP}) onto itself. This shows that the particle density of
the dual dynamic percolation can be identified with the field of the debris of
the primal dynamic percolation process. Relation (\ref{DualTr_2}) leads to the
identity $\beta-\nu z=\tilde{\beta}$ between the exponents defined in
(\ref{GenSkal}), and shows that one has to identify $\tilde{\beta}=\beta_{P}$,
where $\beta_{P}$ is the order parameter exponent defined in static
percolation theory. It is easy to see that the mean field result (\ref{P(t)})
for the survival probability is applicable above six spatial dimensions also
in the case of dynamic isotropic percolation. For logarithmic corrections in
the upper critical dimension $d_{c}=6$ see \cite{JaSt03}. Whereas the formulas
(\ref{LimP}, \ref{LimP_t}) are correct in general, one now has
\begin{eqnarray}
P_{\infty} =\alpha^{2}\lim_{h\rightarrow+0}\langle m(\mathbf{0} 
,\infty)\rme^{\lambda h\dot{\widetilde{\mathcal{N}}}(0)}\rangle
=\alpha^{2}\rho_{st}\propto\left\vert \tau\right\vert ^{\tilde{\beta}}.
\end{eqnarray}
instead of equation (\ref{RelPr}) as a consequence of the symmetry (\ref{DualTr_2}).
Here $\rho_{st}$ is the stationary homogeneous density of the debris in the
active percolating phase.

\subsection{Tricritical dynamic isotropic percolation}

As a last example we consider briefly the tricritical variant of dynamic
isotropic percolation (TdIP) \cite{JaMu-u,JaMuSt04}. TdIP as a tricritical
point followed by a first order transition can be founded on a generalization
of the GEP. One introduces in the reaction scheme of the GEP weakened
individuals as a further species. This species is weakened by an infected
neighbor but not immune. Hence, a second infection in its neighborhood makes
this weakened individuals more easy susceptible. In the consequence, this
property introduces an instability that may lead to a discontinuous epidemic
transition. The response functional (\ref{J_DIP}) is extended to
\begin{eqnarray}
\fl \qquad \mathcal{J}_{TdIP}[\tilde{n},n] =\int \rmd t\,\rmd^{d}x\,\biggl\{\lambda\tilde
{n}\Big(\lambda^{-1}\frac{\partial}{\partial t}+(\tau-\nabla^{2})+fm+\frac{g^{\prime}}{2}m^{2}-\frac{g^{\prime\prime}}{2}\tilde
{n}\Big)n\biggr\}.
\end{eqnarray}
The tricritical point is given in mean field theory by $\tau=f=0$. The upper
critical dimension is $d=d_{c}=5$. The symmetry (\ref{DualTr_2}) is lost.
Hence, we find four independent critical exponents. In a $\varepsilon
$-expansion with $\varepsilon=5-d$ we obtained
\begin{equation}
\beta=\frac{3}{2}-\frac{74}{225}\varepsilon+O(\varepsilon^{2}),\quad
\beta^{\prime}=1-\frac{2}{45}\varepsilon+O(\varepsilon^{2}).
\end{equation}
The mean field result (\ref{P(t)}) for the survival probability is applicable
above five spatial dimensions also in this case. For logarithmic corrections
in the upper critical dimension $d_{c}=5$ see \cite{JaMuSt04}

\section{Epilogue}

Survival and percolation probabilities are most important quantities in the
theory and in the application of spreading. In the past, one has missed
calculable expressions in the field theory of this phenomena. We have seen
that it is possible to derive such expressions, and that their calculations
are feasible in perturbation expansions. We have presented the results in a
simple mean field approximation. Extension to higher order of perturbation
series are possible without difficulties. We have presented results in
\cite{JaSt04,JaSt03,JaMuSt04}.

The outstanding role of the form of the absorptive noise is stressed. So
called multiplicative noise would lead to trivial non-decaying survival
probabilities. In the last chapter we have shortly considered some fundamental
growth models equipped with absorptive noise that are representations of known
universality classes of spreading phenomena. All considerations are applicable
also to models with long-range spreading \cite{JaOeWiHi99}

\ack I thank Georg Foltin and Olaf Stenull for a critical reading of the
manuscript. This work has been supported by the Deutsche
Forschungsgemeinschaft via the Sonderforschungsbereich 237 \textquotedblleft
Unordnung und gro{\ss }e Fluktuationen\textquotedblright.

\appendix

\section*{Appendix: Renormalized field theory of compact directed percolation}
\setcounter{section}{1}

It is well known that CDP offers many exactly calculable properties. The
response functional of the CDP process with bias (\ref{J_VM}) equipped with a
further relevant term to describe the crossover to DP is
\begin{eqnarray}
\fl \mathcal{J}[\tilde{n},n]  =\int \rmd t\,\rmd^{d}x\,\Big\{\lambda\tilde
{n}\Big(\lambda^{-1}\frac{\partial n}{\partial t}-\nabla^{2}n+(\tau
+r)n(1-n)+rn^{2}-\frac{g}{2}\tilde{n}n(1-n)\Big)\Bigr\}\nonumber\\
\fl \qquad\qquad =\int \rmd t\,\rmd^{d}x\,\Bigl\{\lambda\tilde{n}\Big(\lambda^{-1}\frac{\partial
n}{\partial t}-\nabla^{2}n+rn+\tau n(1-n)-\frac{g}{2}\tilde{n}n(1-n)\Big)\Bigr\}. \label{J_CDP1} 
\end{eqnarray}
The new coupling term with $r\geq0$ suppresses now the density of the active
particles in the formerly compact filled regions in contrast to the other
terms. The deterministic term proportional to $\tau$ and the noise
proportional to $g$ are effectively present only at the interface between
empty and filled regions \cite{DCCH01}. These interfaces are broken up if
$r>0$ and the process crosses over to DP. Note that in a mean field picture
the transition line between the absorbing and the active phase is given by
$\tau_{c}=-r$.

The renormalized field theory of the critical voter model below $d=d_{c}=2$,
that is (\ref{J_CDP1}) with $\tau=r=0$, is usually developed by mapping the
model to the problem of annihilating random walks. The mapping consists of two
steps. At first a duality transformation is applied, then the diagrams
contributing to the relevant vertex functions \cite{Pe86,Le94} are summed up
exactly. This can be done also directly for the field theory defined by the
functional (\ref{J_CDP1}). Including the parameter $r$ in the propagator
$G(t,q,r)=\theta(t)\,\exp\bigl(-\lambda(q^{2}+r)t\bigr)$, we find for the
primitively diverging vertex functions $\Gamma_{\tilde{k},k}$, where
$\tilde{k}$ and $k$ denote the numbers of the amputated $\tilde{n}$- and
$n$-legs of the irreducible diagrams, exactly
 \begin{eqnarray} \label{vert_VM}
\Gamma_{2,2}  &  =-2\Gamma_{2,1}=2\lambda gD,\\
\Gamma_{1,2}  &  =-2\Gamma_{1,1}+2(i\omega+\lambda q^{2}+\lambda
r)=-2\lambda\tau D.
\end{eqnarray}
Here,
\begin{equation}
D=\frac{1}{1+gI}, \label{B_C} 
\end{equation}
is the correction that arises from the summation of all contributing diagrams,
the chains of \textquotedblleft bubbles\textquotedblright. The single bubble
integral is given by
\begin{eqnarray}
I(\omega,q,r)  &  =\lambda\int_{0}^{\infty}\rmd t\,\rme^{-i\omega t}\int                     _{p}G(t,\frac{q}{2}+p,r)G(t,\frac{q}{2}-p,r)\nonumber\\
&  =\frac{\Gamma(1+\varepsilon/2)}{\varepsilon(4\pi)^{d/2}}g\Big(r+\frac
{q^{2}}{4}+\frac{i\omega}{2\lambda}\Big)^{-\varepsilon/2}, \label{B_I} 
\end{eqnarray}
where $\varepsilon=2-d$. We renormalize $g$ by $g\rightarrow\mathring{g}=Zg$,
and choose the renormalization point RP by $\bigl(r+q^{2}/4+i\omega
/2\lambda\bigr)=\mu^{2}$, where $\mu$ is the usual convenient inverse length
scale. Furthermore, we define
\begin{equation}
\left.  \mathring{g}D\right\vert _{RP}=\frac{\mathring{g}}{1+\mathring
{g}\left.  I\right\vert _{RP}}=:g=A_{\varepsilon}u\mu^{\varepsilon}
\label{R_Def} 
\end{equation}
with $A_{\varepsilon}=\Gamma(1+\varepsilon/2)/(4\pi)^{d/2}$. We obtain finally
the renormalization constant
\begin{equation}
Z=\frac{1}{1-u/\varepsilon} \, . \label{Z} 
\end{equation}
Equations (\ref{vert_VM}) show that $\tau$ renormalizes in the same way as $g$,
whereas $r$ needs no renormalization:
\begin{equation}
\tau\rightarrow\mathring{\tau}=Z\tau,\quad r\rightarrow\mathring{r}=r.
\label{t,r_Ren} 
\end{equation}
With the help of this procedure all $\varepsilon$-poles are eliminated from
the perturbation series.

In the same way one can study the correlation functions with the composite
interface field $\varphi=n(1-n)$. An insertion of the field $\varphi$ requires
a further renormalization. It can be easily seen that this renormalization is
given by $\varphi\rightarrow\mathring{\varphi}=Z^{-1}\varphi$. Thus, all
renormalizations of the response functional (\ref{J_CDP1}) follow from the
renormalization of this interface field. The cumulant of $k$ $n$-fields,
$\tilde{k}$ $\tilde{n}$-fields and $m$ $\varphi$-fields, starting from
uncorrelated initial conditions with $n(\mathbf{x},t=0)=n_{0}$ is denoted by
$G_{k,\tilde{k};m}$. The renormalization group equation (RGE) for the Greens
functions $G_{k,\tilde{k};m}$, reads
\begin{equation}
\Big(\mu\partial_{\mu}+\beta\partial_{u}+\kappa\tau\partial_{\tau}
+m\kappa\Big)G_{k,\tilde{k};m}(\{\mathbf{x},t\},\tau,r,n_{0},u,\lambda,\mu)=0,
\label{RGG} 
\end{equation}
where the RG-functions result from the logarithmic derivative of the
$Z$-factor, $\kappa=-\left.  \partial\ln Z/\partial\ln\mu\right\vert _{0}$,
holding bare parameters constant. We get
\begin{equation}
\kappa=u,\quad\beta=(-\varepsilon+u)u. \label{RG-Fu} 
\end{equation}
At the stable fixed point of the RGE, $u_{\ast}=\varepsilon>0$, we arrive the
scaling form
\begin{eqnarray}
\fl \qquad G_{k,\tilde{k};m}(\{\mathbf{x},t\},\tau,r,n_{0},u_{\ast})=l^{\tilde{k}d+m\varepsilon}G_{k,\tilde{k};m}(\{l\mathbf{x},l^{2}t\},\tau
l^{-d},rl^{-2},n_{0},u_{\ast}). \label{Gr-Skal} 
\end{eqnarray}
All the critical exponents follow from these equations. The DP critical line,
the line between the absorbing and the active phase for $r>0$, results as
$\tau_{c}\propto-r^{d/2}$. From the asymptotic solution of the RGE directly in
two spatial dimensions, one can infer
\begin{eqnarray}
\fl \qquad G_{k,\tilde{k};m}(\{\mathbf{x},t\},\tau,r,n_{0},u)  =l^{2\tilde{k}}
X(l)^{m} G_{k,\tilde{k};m}(\{l\mathbf{x},l^{2}t\},X(l)l^{-2}
\tau,rl^{-2},n_{0},X(l)u),\nonumber\\
X(l)^{-1}   =1-u\ln l. \label{LogKorr1} 
\end{eqnarray}
This general scaling relation determines all the logarithmic corrections known
for CDP in $d=2$. One gets, e.g., for the mean interface density at the
critical point $\tau=0$ and with a non-universal time constant $t_{0}$,
asymptotically
\begin{equation}
\langle\varphi(\mathbf{x},t)\rangle\propto n_{0}(1-n_{0})(\ln t/t_{0}
)^{-1}+O((\ln t/t_{0})^{-2}),
\end{equation}
whereas $\langle n(\mathbf{x},t)\rangle=n_{0}=\mathit{const}$, a relation well
known from the exact solution of the two-dimensional lattice voter model
\cite{SS88,Kr92}.

\end{document}